\documentclass[
    prd,
    reprint,
    amsmath,
    amssymb,
    floats,
    superscriptaddress,
    nofootinbib,
    showpacs
]{revtex4-2}

\usepackage[utf8]{inputenc} 
\usepackage{graphicx,graphics,xcolor,overpic,mathtools}
\usepackage{amsthm,amsmath,amssymb}
\usepackage[colorlinks]{hyperref}
\usepackage{textcomp}
\definecolor{coolblack}{rgb}{0.0, 0.18, 0.39}
\hypersetup{
    colorlinks = false,
   }
\PassOptionsToPackage{normalem}{ulem}

\usepackage{ulem} 
\usepackage{appendix}
\usepackage{bm}
\usepackage{url}
\usepackage{physics}
\usepackage[makeroom]{cancel}
\usepackage{cleveref}
\usepackage{todonotes}
\usepackage{tensor}
\usepackage{mathrsfs}
\usepackage{slashed}
\usepackage[mode=buildnew]{standalone}

\newcommand{\comment}[1]{}

\usepackage{pgfplots}

\usepackage{diagbox}
\usepackage{stackengine,amssymb,graphicx}

\NewDocumentCommand{\evat}{sO{\bigg}mm}{%
  \IfBooleanTF{#1}
   {\mleft. #3 \mright|_{#4}}
   {#3#2|_{#4}}%
}
\usepackage{enumerate}
\definecolor{azure}{rgb}{0.0, 0.5, 1.0}

\begin{document}

%\title[]{Galactic halos and subhalos composed by bosonic dark matter}
\title[]{Reproducing galactic rotation curves with a two-component bosonic dark matter model}

\author{Jorge Castelo Mourelle}

\affiliation{%
Departamento de Astronom\'ia y Astrof\'isica, Universitat de Val\`encia,
Dr. Moliner 50, 46100 Burjassot (Val\`encia), Spain
}%

\author{Nicolas Sanchis-Gual}

\affiliation{%
Departamento de Astronom\'ia y Astrof\'isica, Universitat de Val\`encia,
Dr. Moliner 50, 46100 Burjassot (Val\`encia), Spain
}%

\author{Jos\'e A. Font}

\affiliation{%
Departamento de Astronom\'ia y Astrof\'isica, Universitat de Val\`encia,
Dr. Moliner 50, 46100 Burjassot (Val\`encia), Spain
}%
\affiliation{%
Observatori Astronòmic, Universitat de València, C/ Catedrático José Beltrán 2, 46980, Paterna (València), Spain
}%

\date[ Date: ]{\today}
\begin{abstract}
Bosonic stars — hypothetical astrophysical entities — are generally categorized into two primary classes based on the nature of their constituent particles: Einstein-Klein-Gordon stars, made up of massive scalar bosons, and Proca stars, 
their vector ``cousins''. Depending on the boson masses and field frequencies, these objects may exhibit properties of diffuse, massive structures, with sizes comparable to or even exceeding those of galaxies. This concept has inspired the bosonic dark matter halo hypothesis, providing a theoretical framework to effectively model the dark matter content of galactic halos.
In this paper we build on our previous work to explore the possibility of using vector and scalar bosons to model the components of galactic dark matter halos and subhalos in order to reproduce the observed rotation curves of galaxies.
By employing diverse combinations of those bosonic dark matter models in conjunction with observable data for a sample of galaxies, we show that our two-component dark matter approach notably improves the agreement between observations and theoretical predictions with respect to our previous investigation. Our framework may shed new light on the enduring mystery surrounding the apparent matter deficit observed in dwarf and spiral galaxies. 
\end{abstract}

\maketitle

%\begin{quote}
 
%\end{quote}

%\begin{minipage}{\textwidth}
%\tableofcontents
%\end{minipage}

%%%%%%%%%%%%%%%%%%%%%%%%%%%%%%%%%%%%%%%%%%%%%%%%%%%%%%%%%%%%%%
\section{\textbf{Introduction.}}
%%%%%%%%%%%%%%%%%%%%%%%%%%%%%%%%%%%%%%%%%%%%%%%%%%%%%%%%%%%%%%

Observations of galaxy and galaxy cluster rotation curves reveal a striking deviation from classical expectations. Instead of exhibiting a Keplerian decline, the measured velocities remain unexpectedly flat, extending far beyond the visible boundaries of galaxies \cite{peebels,doi:10.1126/science.220.4604.1339,Persic:1995ru,Primack:2001ia}. This persistent flatness, commonly known as the Rotation Curves (RC) problem, constitutes a critical argument in favor of non-baryonic dark matter (DM). Various theories have been proposed to explain these anomalies. Some authors have suggested modifications to Newtonian gravity \cite{1983ApJ...270..365M}, while others advocate for the existence of invisible, non-interacting DM~\cite{Chardin:1999jj,Baltz:1997dw,Beylin:2020bsz}. The early observations of the Coma cluster \cite{2009GReGr..41..207Z} together with more precise measurements of galaxy RC during the 1970s reinforced the DM hypothesis \cite{1970ApJ...159..379R}. Freeman’s model of spherical halos introduced the concept of a linearly increasing mass function, and subsequent studies have mapped DM halos exceeding quantitatively the observable galactic regions \cite{1970ApJ...160..811F,Persic:1995ru,Brainerd:1995da,Lanzetta_1996,Primack:1997av,Zaritsky:1996ch}.

The possibility that galactic halos could be composed of bosonic DM  has also been investigated in several works \cite{Schunck:1998nq,Urena-Lopez:2010zva,Broadhurst:2019fsl,Chen:2020cef,Annulli:2020ilw,Amruth:2023xqj,Pozo:2023zmx}. In particular, models incorporating an ultralight axion-like particle have attracted much attention, as they naturally give rise to DM halos modeled as Newtonian Bose-Einstein Condensates \cite{Luu:2018afg,deMartino:2018krg,Emami:2018rxq,2024arXiv240104735K,Sakharov:2021dim}. The Scalar Field Dark Matter model, which is consistent with the $\Lambda$CDM paradigm, predicts large-scale phenomena that align with linear-order perturbations \cite{PhysRevD.68.024023,PhysRevD.35.3640,Sahni:1999qe,Matos:2000ss,Matos:2008ag,Urena-Lopez:2010zva}. By employing ground state solutions of the Schrödinger-Poisson (SP) system—the only stable configuration where all bosonic particles reside in the lowest energy state—these models successfully reproduce the observed RC. Stability analyses further confirm that while the ground state is robust against gravitational perturbations, excited state configurations remain inherently unstable \cite{Sin:1992bg,Ji:1994xh,PhysRevD.69.124033,Lee:1995af,Arbey:2003sj,Guzman:2006yc,Ruffini:1969qy,Guzman:2004wj,Schunck:2003kk}.

Models in which a bosonic field minimally coupled to gravity acts as a source of DM are directly linked with bosonic stars.  
These can be primarily classified into scalar boson stars (BS) and Proca stars (PS), which are localized, regular, horizonless solutions modeled by massive, free or self-interacting, complex scalar and vector fields bound by gravity, respectively  \cite{PhysRev.172.1331,PhysRev.187.1767,PhysRev.148.1269,Schunck:2003kk,Brito:2015pxa,Herdeiro:2023wqf}. Recent advances have expanded this framework to include Proca-Higgs stars (PHS), where complex vectors interact with real scalars to yield richer dynamics \cite{Herdeiro:2023lze,Brito:2024biy}. Moreover, investigations into multi-field configurations have led to the development of multi-state boson stars and $\ell$-bosonic stars, thereby broadening the spectrum of viable models \cite{Urena-Lopez:2010zva,Alcubierre:2018ahf,jaramillo2020dynamical,Sanchis-Gual:2021edp,Lazarte:2024jyr} (see~\cite{Liebling:2023,bezares2025exotic} for recent reviews).

The theoretical feasibility of bosonic stars has been extensively examined, particularly regarding their formation mechanisms, stability conditions, and dynamical behavior~\cite{PhysRev.97.511,PhysRev.172.1331,PhysRev.187.1767,PhysRev.148.1269,Lai:2004fw,Schunck:2003kk,Liebling:2023,bezares2025exotic}. While initially conceptually conceived as static and spherically symmetric objects, modern studies now routinely explore rotating configurations that manifest as axisymmetric, spinning solutions in both scalar and vector forms \cite{Schunck:1996he,Yoshida:1997qf,Brito:2015pxa,Herdeiro:2016tmi,Herdeiro:2019mbz,Sanchis-Gual:2019ljs}. Their versatility in emulating a range of astrophysical objects—including neutron stars, black holes, and intermediate-mass bodies—renders them a powerful tool in astrophysical modeling, allowing to explore the effect of purely gravitational entities~\cite{Schunck:1998nq,Adam:2010rrj,PhysRevD.80.084023,Herdeiro:2021lwl,Rosa:2022tfv,Rosa:2023qcv,Sengo:2024pwk}.

\begin{figure}[t]
\centering
\includegraphics[width=0.47\textwidth]{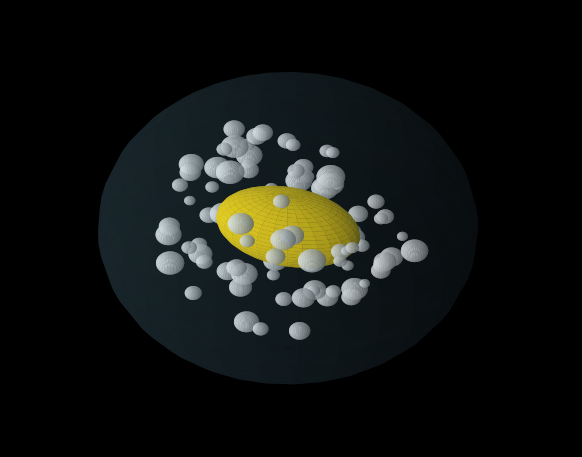}
\caption{Schematic illustration of the model.  }
\label{model_pic}
\end{figure}

This paper explores the modeling of galactic DM halos using bosonic fields, extending the study initiated in~\cite{Mourelle:2024dlt} and addressing two open issues identified in that work. First, the previous analysis changed the properties of the vector field for each galaxy instead of treating it as a single DM candidate; and second, it introduced an additional dark component without a clear physical justification. As in~\cite{Mourelle:2024dlt} we employ bosonic vector fields coupled to gravity to represent the primary galactic halo, thus enhancing RC fits when combined with ordinary matter contributions. A significant modification in the present work, which addresses the first open issue, is that we now fix the vector boson mass to a specific scale relevant to the problem, allowing only for the field frequency to vary, thereby identifying constraints on this parameter. Regarding the second issue, in~\cite{Mourelle:2024dlt} the extra component was introduced through an \textit{ad-hoc} mathematical adjustment, lacking physical motivation and coherence across the configurations. Here, we provide a physically meaningful explanation by employing a subhalo model consisting of a scalar bosonic field coupled to gravity to represent intermediate galactic structures. An illustration of our model is depicted in Fig.~\ref{model_pic}. It consists of the following components: the luminous matter contribution, represented by a yellow ellipsoid (the galaxy); a quasi-spherical main halo formed by rotating vector bosonic matter extending beyond the galaxy, shown in dark grey; and a set of spherical subhalos modeled as scalar boson star-like structures, depicted in light grey. As we show in this paper, this model significantly improves the RC fits presented in \cite{Mourelle:2024dlt}, in addition to provide a physically justified framework for them.

Multi-component bosonic dark matter scenarios, involving both ultra-light scalars and ultra-light vectors, can naturally arise in well-motivated extensions of the Standard Model. For instance, string theory compactifications often predict axion-like particles and dark photons as part of the low-energy spectrum~\cite{Arvanitaki:2009fg}. Similarly, dark sector models with a broken $\mathrm{U}(1)$ gauge symmetry can accommodate both light vector bosons and ultra-light scalars, such as axions or relaxions, with stability ensured by symmetries or suppressed couplings~\cite{Kaneta:2017wfh, Adams:2020uco, Chu:2019rok}. Minimal scalar and gauge extensions of the Standard Model can also lead to coexisting bosonic relics produced through non-thermal mechanisms and stable on cosmological timescales~\cite{Freitas:2021cfi}.

The structure of the paper is as follows: In Section~\ref{setup} we introduce the theoretical framework for the bosonic fields we employ in our model, along with key clarifications regarding the rescalings and physical units used. 
%In Section~\ref{geodesicssect} we present the geodesic orbits from which the rotation curves are derived. 
Section~\ref{models} details how the individual contributions to the RC are obtained from both luminous and DM systems. In particular, this section discusses the model for DM subhalos under the assumption that they could be formed by a distribution of boson stars.  
Next, in Section~\ref{results} we compare our model predictions with observational data, using the same sample of galaxies employed in~\cite{Mourelle:2024dlt}. This section also provides a quantitative assessment of our new model against the fits reported in~\cite{Mourelle:2024dlt}. A discussion of this study is presented in Section~\ref{conclusions} along with our conclusions. Additional information on the equations of motion for the scalar and vector bosonic models is provided in Appendix~\ref{AppendixA}.

%%%%%%%%%%%%%%%%%%%%%%%%%%%%%%%%%%%%%%%%%%%%%%%%%%%%%%%%%%%%%%
\section{Bosonic fields coupled to gravity} 
\label{setup}
%%%%%%%%%%%%%%%%%%%%%%%%%%%%%%%%%%%%%%%%%%%%%%%%%%%%%%%%%%%%%%

\subsection{Formalism}

The two main families of bosonic stars are described within the Einstein-Klein-Gordon (EKG) and Einstein-Proca (EP) systems, where a massive complex scalar field $\Phi$ or a massive complex vector field $A$, respectively, are minimally coupled to Einstein's gravity. The overall action is given by 
\begin{equation} 
\mathcal{S}{(\Phi,A)} = \int \left(\frac{1}{16\pi } R + \mathcal{L}_{(\Phi,A)} \right) \sqrt{-g} d^4x, 
\label{action} 
\end{equation}
with $R$ denoting the Ricci scalar, $g$ the determinant of the metric, and $\mathcal{L}_{(\Phi,A)}$ representing the Lagrangian densities for the scalar or vector fields. The general metric for a static spherical symmetric system, in Schwarzschild-like coordinates, can be written in terms of two real metric functions $\gamma(r)$ and $\lambda(r)$ as follows,
 \begin{equation}
    ds^{2}=-e^{\gamma\left(r\right)}dt^{2}+e^{\lambda\left(r\right)}dr^{2}+r^{2}\left[d\theta^{2}+\sin^{2}\left(\theta\right)d\psi^{2}\right].
    \label{metricBSsch}
 \end{equation}
The coordinate  $r$ represents an areal radius, meaning that spheres of constant $r$ have a surface area of $4\pi r^2$.

For scalar boson stars, the Lagrangian governing the complex field dynamics reads,
\begin{equation}
\mathcal{L}_{\Phi} = -\frac{1}{2} \left[g^{\alpha\beta} \nabla_{\alpha} \Phi^* \nabla_{\beta} \Phi + V\left(|\Phi|^2\right) \right], \label{lagrangianscalar} 
\end{equation}
where the potential $V\left(|\Phi|^2\right)$, preserving a global $U(1)$ symmetry, includes the quadratic mass term $\mu_{\rm S}^2|\Phi|^2$ along with self-interaction terms that determine the star's properties. Varying the action (\ref{action}) with this Lagrangian yields the EKG equations:
\begin{equation}
\begin{split}
&R_{\alpha\beta}-\frac{1}{2}Rg_{\alpha\beta}=8\pi T^{\Phi}_{\alpha\beta}, \\ &g^{\alpha\beta}\nabla_{\alpha}\nabla_{\beta}\Phi=\frac{dV}{d|\Phi|^2}\Phi, \label{kg} \end{split} \end{equation}
where the scalar field stress-energy tensor is given by 
\begin{equation}
\begin{split}
T_{\alpha\beta}^{\Phi}=& \nabla_{\alpha}\Phi^*\nabla_{\beta}\Phi+\nabla_{\beta}\Phi^*\nabla_{\alpha}\Phi\\ &- g_{\alpha\beta}\left[g^{\mu\nu}\left(\nabla_{\mu}\Phi^*\nabla_{\nu}\Phi+\nabla_{\nu}\Phi^*\nabla_{\mu}\Phi\right)+V\left(|\Phi|^2\right)\right]. \label{stress} \end{split} \end{equation}
To ensure staticity and spherical symmetry, the scalar field has the form \cite{Liebling:2012fv},
\begin{equation}
    \Phi(t,r)=\Phi_0(r)e^{iwt},
    \label{field_static_BS}
\end{equation}
where $w$ is the internal field frequency, and $\Phi_0$ is a real scalar taking the role of the star's profile. 
A complete study of the topic is shown in \cite{CasteloMourelle:2025ujn}.

Correspondingly, the Lagrangian density for vector boson stars is
\begin{equation}
\mathcal{L}_A=-\frac{1}{4}F_{\alpha\beta}F^{*\alpha\beta}-\frac{1}{2}\mu_{\rm V}^2A_{\alpha}A^{*\alpha}, 
\end{equation}
where $F_{\alpha\beta}=\partial_{\alpha}A_{\beta}-\partial_{\beta}A_{\alpha}$, and the Euler-Lagrange equations enforce the Lorenz condition $\nabla_{\alpha}A^{\alpha}=0$. 

To incorporate rotation, we adopt a stationary and axisymmetric metric ansatz \cite{Herdeiro:2015gia,PhysRevD.55.6081}:
\begin{equation} 
\begin{split} ds^2 =& -e^{2\nu}dt^2 + e^{2\beta}r^2\sin^2\theta\left(d\psi-\frac{W}{r}dt\right)^2 \\\ &+ e^{2\alpha}(dr^2+r^2d\theta^2), \label{axmetric} \end{split} \end{equation} 
where the metric functions $\nu$, $\alpha$, $\beta$, and $W$ depend only on $r$ and $\theta$.

The corresponding field equations, describing the EP system, are 
\begin{equation} 
\begin{split} 
&R_{\alpha\beta}-\frac{1}{2}Rg_{\alpha\beta}=8\pi T^{A}_{\alpha\beta}, \\ &\nabla_{\alpha} F^{\alpha\beta}=\mu_{\rm V}^2A^{\beta}, \label{pr} 
\end{split}
\end{equation}
with the stress-energy tensor for the vector field given by 
\begin{equation} 
\begin{split}
T^{A}_{\alpha\beta}=& -F_{\sigma(\alpha}F_{\beta)}^{*\sigma}-\frac{1}{4}g_{\alpha\beta}F_{\mu\nu}F^{*\mu\nu}\\ &+\mu_{\rm V}^2\left[A_{(\alpha}A^*_{\beta)}-\frac{1}{2}g_{\alpha\beta}A_{\sigma}A^{*\sigma}\right]. \end{split} \end{equation}

The vector field is assumed to adopt the form \cite{Brito:2015pxa} 
\begin{equation}
A = e^{i(n\psi-wt)}\left(i V dt+\frac{H_1}{r}dr+H_2 d\theta+iH_3\sin\theta d\psi \right), 
\end{equation}
where the functions $V$, $H_1$, $H_2$, and $H_3$ depend on $r$ and $\theta$, and $n$ is again the azimuthal harmonic index. In our analysis, we restrict to the case $n=1$ for the EP models, since larger values result in rotation curves with velocities exceeding observational limits.

%%%%%%%%%%%%%%%%%%%%%%%%%%%%%%%%%%%%%%%%%%%%%%%%%%%%%%%%
\subsection{Numerics, physical units and rescaling}
\label{rescalingsect}
%%%%%%%%%%%%%%%%%%%%%%%%%%%%%%%%%%%%%%%%%%%%%%%%%%%%%%%%

The static boson stars considered as subhalo structures are governed by the EKG system. They are modeled by ODEs subject to boundary conditions, forming an eigenvalue problem (see Appendix~\ref{AppendixA} for technical details). We solve this set of ODEs using a Python-based shooting method, adjusting either the central field value or the internal frequency to satisfy the asymptotic conditions at infinity. After rescaling all quantities by the boson mass (see below), the only free parameters are the frequency, the initial field value, and possible self-interaction couplings. We integrate the equations using \texttt{solve$\_$ivp}, iteratively refining the shooting parameter until the boundary conditions are met within a predefined tolerance. Moreover, to ensure physical observables and the correct asymptotically flat behavior, we rescale the metric using the value of $g_{rr}$ at infinity, correcting both $g_{tt}$ and the frequency via 
\begin{eqnarray}
w_{\text{real}} = \frac{w_{\text{shoot}}}{\sqrt{g_{tt}|_\infty g_{rr}|_\infty}}.
\end{eqnarray}

To build the main halo component we need to integrate the EP system. This system consists of eight equations, namely four for the metric functions and four for the field functions (see Appendix~\ref{AppendixA} for details). The numerical integration is performed using a normalization of the radial distance and angular frequency by the mass $\mu_{\rm V}$ of the Proca field, through the transformations shown below. Therefore, we remove the explicit dependence on $\mu_{\rm V}$ in the field equations, allowing for numerical solutions to be obtained in units of $\mu_{\rm V}$. A specific physical value for the bosonic mass can be reinstated when returning to dimensional units for the observables, with the rescaling procedure detailed below. 

Specifically, we solve the EP system using the FIDISOL/CADSOL package \cite{schonauer1989efficient,schonauer2001we}, a Newton-Raphson-based solver that features a flexible grid and order of accuracy, and provides an error estimate for each unknown function. To deal with the infinite range of the radial coordinate, we compactify it via the transformation $x \equiv r/(c + r)$, mapping $r \in [0, \infty)$ to a finite interval $x \in [0, 1]$. The equations are discretized on a $(265 \times 35)$ $(x, \theta)$ grid, with $\theta \in [0, \pi/2]$. To ensure sufficient resolution in the far-field region, especially for certain frequency ranges, we adjust the compactification parameter to $c = 1.6$. We address the reader to \cite{CasteloMourelle:2025ujn} for further details on the numerical methodologies we use to solve the EKG and EP systems.

In our numerical calculations, as mentioned above, all quantities are rendered dimensionless by rescaling with the boson mass, assuming $c=G=\hbar=1$. The dimensionless variables (denoted by a tilde) are defined as
\begin{eqnarray}
\tilde{r} = r\mu_{\rm (S,V)},\quad \tilde{t} = t\mu_{\rm (S,V)},\quad \tilde{\omega} = \frac{\omega}{\mu_{\rm (S,V)}}.
\label{rescal}
\end{eqnarray}
Since $\mu_{\rm (S,V)}$ has dimensions of inverse length (related to the reduced Compton wavelength of the boson), the physical mass $m$ is expressed as
\begin{eqnarray}
m = \frac{\hbar \mu_{\rm (S,V)}}{c}.
\end{eqnarray}
In astrophysical applications, masses are often given in energy units (eV), using the conversion
$
1\,\text{eV}/c^2 = 1.783\times10^{-36}\,\text{kg},
$
and the reduced Compton wavelength is computed using $\hbar = 1.054\times10^{-34}\,\text{J.s}$ and $c = 2.998\times10^8\,\text{m/s}$. Distances in astrophysics are usually expressed in kiloparsecs (kpc), with the conversion
$
1\,\text{kpc} = 3.086\times10^{19}\,\text{m}.
$

The scaling of the total mass of the bosonic star goes like
$
\tilde{M} = M\mu_{\rm (S,V)},
$
or equivalently,
\begin{eqnarray}
\tilde{M} = \frac{M\mu_{\rm (S,V)} L_{\rm Planck}}{M_{\rm Planck}} = \frac{Mm}{M_{\rm Planck}^2},
\end{eqnarray}
where the Planck mass is defined by
$
M_{\rm Planck} = \sqrt{{\hbar c}/{G}} \approx 2.1764\times10^{-8}\,\text{kg}, 
$
the Planck length is $L_{\rm Planck}=\sqrt{\hbar G/c^3}\approx 1.616\cdot 10^{-35}\,\text{m}$
, and a solar mass is
$
1\,M_\odot = 1.9884\times10^{30}\,\text{kg}.
$
Finally, the rotation curves computed in our simulations yield velocities in units of $c$.
A comprehensive discussion of these scaling laws can be found in~\cite{Mourelle:2024dlt}.

This unified numerical framework ensures accurate resolution of the EKG and EP models, while providing a clear pathway for converting the dimensionless results into physical observables relevant for astrophysical applications.
 
%%%%%%%%%%%%%%%%%%%%%%%%%%%%%%%%%%%%%%%%%%%%%%%%%%%%
%\section{Matter description and observable curves}
\section{Contributions to galactic rotation curves}
\label{models}
%%%%%%%%%%%%%%%%%%%%%%%%%%%%%%%%%%%%%%%%%%%%%%%%%%%%

A typical galaxy comprises several fundamental components: the bulge, the disk, formed by a stellar component and a gaseous one, and the halo \cite{Lelli:2016zqa,binney2008galactic}. The central bulge is a dense, spherical region predominantly hosting older stars. Surrounding the bulge, the stellar disk forms a flattened, rotating structure containing most of the galaxy's stars, including younger and brighter populations. The gas component of a galactic disk is primarily composed of hydrogen gas essential for star formation. Surrounding these luminous components is the halo, a more extended, roughly spherical region that includes older stars, globular clusters, and DM \cite{DeLaurentis:2022nrv}, significantly dominating the galaxy's total mass and gravitational influence \cite{binney2008galactic,donato2009constant}. In this study we model the luminous galaxy components (bulge and disk) using the same methodology as in \cite{Mourelle:2024dlt}. For the halo constituents, however, we consider here vector bosons as the primary DM halo components, supplemented by scalar bosonic subhalo structures.

In our model, both the host halo and the subhalo population are composed of bosonic fields: the main halo is modelled as a dilute distribution of vector bosons, while the subhalos are compact scalar configurations. Both components are assumed to form in the early Universe through gravitational condensation \cite{Kolb:1993zz,Levkov:2018kau,Seidel:1993zk}. The scalar subhalos, low in mass and solitonic in nature, are highly stable against dynamical friction~\cite{Hui:2016ltb,bar2019relaxation} and tidal stripping~\cite{Schive:2014dra,Marsh:2018zyw}, especially when embedded in a coherent vector boson halo. Their survival over cosmological timescales is supported by their compactness and the suppressed drag expected from a smooth or condensate-like host halo. This makes the coexistence of both components, and the persistence of substructure to the present day, a natural outcome of this two-component bosonic dark matter scenario~\cite{Schwabe:2016rze,Freitas:2021cfi}.

Given the various components, galactic RC can be analyzed by decomposing them into distinct velocity contributions, as we do next, each of them modeled independently based on measurable luminosities.  %The models employed for these luminous and dark matter components are detailed in what follows.
The total velocity is hence given by
\begin{equation}
    v_{\rm T}^2(r)=v_{\rm disk}^2(r)+v_{\rm gas}^2(r)+v_{\rm bulge}^2(r)+v_{\rm Proca}^2(r)+v_{\rm SH}^2(r)\,,
\end{equation}
where each of the terms on the r.h.s.~corresponds to the contribution from the stellar disk, from the gas disk, from the bulge, from the Proca DM halo, and from the distrubution of the scalar DM subhalos.

%\jcm{The reason for having a quadratic addition law is due to the linear nature of the gravitational potential. In a spherical or axially symmetric system, the necessary centripetal acceleration required for maintain a circular orbit at a given radius $r$ is given by the derivative of the gravitational potential, and as it is linear, the orbit is also linear in the square of the velocities:
%$v(r)^2\sim d\Phi/dr$ and $\Phi(r)=\Phi_1+\Phi_2+...$}

%%%%%%%%%%%%%%%%%%%%%%%%%%%%%%%%%%%%%%%%%%%%%%%%%%%%%%%%%%%%%%
\subsection{Disk}
%%%%%%%%%%%%%%%%%%%%%%%%%%%%%%%%%%%%%%%%%%%%%%%%%%%%%%%%%%%%%%

We separate the disk contribution into two distinct velocity components, stellar matter and neutral hydrogen (HI) gas. For the stellar disk, we adopt Freeman's thin disk approximation \cite{1970ApJ...160..811F,galacticDyn},
\begin{equation}
v_{\rm disk}^2(r)=4\pi G\Sigma_0 R_{\rm disk} y^2[I_0(y)K_0(y)-I_1(y)K_1(y)],
\end{equation}
where $\Sigma_0=M_{\rm disk}/(2\pi R_{\rm disk}^2)$, $M_{\rm disk}$ is derived from total luminosity and mass-to-luminosity ratio, $R_{\rm disk}$ is the empirically determined disk scale radius, $y=r/(2R_{\rm disk})$, and $I_n(y)$, $K_n(y)$ are modified Bessel functions.

For the gas disk, we use an exponential density profile, commonly employed in galactic modeling \cite{binney2008galactic,bosma1981distribution,begeman1989hi,deblok2008high}:
\begin{equation}
\Sigma_{\text{HI}}(r) = \Sigma_{0,\text{HI}} \exp\left(-\frac{r}{R_{\rm disk}}\right),
\end{equation}
where $\Sigma_{0,\text{HI}}$ is the galaxy-specific central HI surface density. The corresponding gas velocity at radius $r$ is given by
\begin{equation}
v_{\text{gas}}(r) = \sqrt{ \frac{G M_{\text{gas}}(r)}{r}}, \quad M_{\text{gas}}(r) = \int_0^r 2 \pi r' \Sigma_{\text{HI}}(r') dr'.
\end{equation}

%%%%%%%%%%%%%%%%%%%%%%%%%%%%%%%%%%%%%%%%%%%%%%%%%%%%%%%%%%%%%%
\subsection{Bulge}
%%%%%%%%%%%%%%%%%%%%%%%%%%%%%%%%%%%%%%%%%%%%%%%%%%%%%%%%%%%%%%

For the bulge, we employ the Sérsic profile, a generalized version of the de Vaucouleurs function,
\begin{equation}
I(r)=I_e e^{-b_n\left[\left(r/R_{eb}\right)^{1/n}-1\right]},
\end{equation}
where $b_n=7.6692$, $n=4$, $R_{eb}$ is the bulge's effective radius, and
\begin{equation}
I_e=\frac{L_{\text{bulge}} b_n^{2n}}{2\pi n e^{b_n}\Gamma(2 n)R_{eb}^2},
\end{equation}
with $L_{\text{bulge}}$ representing the bulge luminosity fraction \cite{Sofue:2008wt,Caon:1993wb}, and $\Gamma(x)$ denoting the Euler gamma function. Since velocities from this profile lack a simple analytical form, we numerically integrate to obtain the enclosed mass:
\begin{equation}
M_{\rm bulge}(r)=\int_0^r 2\pi r' I(r')dr'
\rightarrow
v_{\rm bulge}(r)=\sqrt{\frac{GM_{\rm bulge}(r)}{r}}.
\end{equation}

%%%%%%%%%%%%%%%%%%%%%%%%%%%%%%%%%%%%%%%%
%\subsection{Proca stars as DM}
\subsection{Primary DM halo: Proca stars}
\label{geodesicssect}
%%%%%%%%%%%%%%%%%%%%%%%%%%%%%%%%%%%%%%%

Following the approach laid out in~\cite{Mourelle:2024dlt}, we model the main halo component using orbital velocities derived from EP systems. The derivation of particle orbits in axisymmetric spacetimes—crucial for studying accretion dynamics around relativistic objects—is detailed in~\cite{Mourelle:2024dlt,Gourgoulhon:2010ju,Grandclement:2014msa,Meliani:2015zta}. In such spacetimes, stable circular orbits are not always permitted and exist only for radii $r > r_{\rm ISCO}$, where $r_{\rm ISCO}$ denotes the innermost stable circular orbit.

The orbital velocity for matter moving in equatorial circular orbits around axisymmetric gravitational sources (with metric given by Eq.~(\ref{axmetric})) is~\cite{Mourelle:2024dlt}
\begin{equation}
\begin{split}
v(r)_\pm = \frac{e^{\beta-\nu}\, r\, \frac{\partial}{\partial r}\left(\frac{W}{r}\right) \pm \sqrt{D}}{2\left(\frac{\partial \beta}{\partial r}+\frac{1}{r}\right)},
\end{split}
\label{velocityhalo}
\end{equation}
where
\begin{equation}
\begin{split}
D = e^{2\beta-2\nu}\, r^2 \left(\frac{\partial}{\partial r}\left(\frac{W}{r}\right)\right)^2 + 4\left(\frac{\partial \beta}{\partial r}+\frac{1}{r}\right)\frac{\partial \nu}{\partial r}.
\end{split}
\label{d}
\end{equation}
Here, the $\pm$ signs correspond to prograde and retrograde orbits, respectively~\cite{Grandclement:2014msa}. The existence of circular orbits is guaranteed when $D \geq 0$.

From the metric functions of a given model, we derive the corresponding orbital velocity profiles around the gravitational source. In the case of EP models, these velocities directly describe the halo contribution to the rotation curves, as the vector boson field is interpreted as a DM component. It is important to note that the characteristic scale of the EP system is set by the boson mass $\mu_{\rm V}$. In this work, we fix this mass to parametrize the spinning EP configurations as galactic halos appropriately. This means that $v(r)_{+}$ in Eq.~(\ref{velocityhalo}) is exactly what we will call $v_{\rm P}(r)$, the velocity component accounting for the vectorial DM halo, when fitting the observational RC.

%Following the previous line, the mass of the EP systems is computed in units of the boson mass. After solving the metric equations, the corresponding mass is extracted via the Komar integral applied to each system's stress-energy tensor \cite{stergioulas1994comparing},
%\begin{equation}
%M_{A} = \int \left[-2T_{t}^{At} + T_{\sigma}^{A\sigma}\right] \sqrt{-g} \,d^3x,
%\label{masss}
%\end{equation}

It is important to emphasize that by considering the non-relativistic limit ($w/\mu_{\rm V} \rightarrow 1$), our framework aligns closely with fuzzy DM models typically applied in such contexts \cite{Painter:2024rnc,Bullock:2017xww,DelPopolo:2016emo}.
Unlike in \cite{Mourelle:2024dlt} we now fix the vector boson mass at $\mu_{\rm V} = 9\times 10^{-26}\, \mathrm{eV}$, setting the halo length scales between $\sim5-50$ kpc. Consequently, the only remaining adjustable parameter is the internal field frequency ($w/\mu_{\rm V}$). Thus, our halo rotation curves take the form $v_{\rm P}(r)=\bar{v}_{\rm P}^{w/\mu_{\rm V}}(r)$, where the ratio $w/\mu_{\rm V}$ varies within the non-relativistic limit, allowing us to choose optimal values for each DM halo candidate.
Additionally, since $\mu_{\rm V}$ is fixed to represent the primary halo without overfitting observational data, we include a secondary dark matter contribution via subhalo structures. 

%%%%%%%%%%%%%%%%%%%%%%%%%%%%%%%%%%%%%%%%%%%%%
\subsection{DM subhalos}
%%%%%%%%%%%%%%%%%%%%%%%%%%%%%%%%%%%%%%%%%%%%

In~\cite{Mourelle:2024dlt} it was shown how using a baseline fit with the EP model and adding a second DM halo improves the fitting of the RC considerably (see, in particular, Section IV.B of~\cite{Mourelle:2024dlt}). The origin of this second DM source is unknown, which somehow points to a limitation of the model. In the model upgrade we report here, an additional component is added to the rotation curve, attributed to a distribution of different scalar bosonic objects, confined to a restricted radial interval. Specifically, the subhalo objects are distributed in the interval
$[r_{\text{inner}},\, r_c],$ so that outside this range (i.e.~for $r < r_{\text{inner}}$ and $r > r_c)$ no subhalo objects are present. The internal distribution is modeled using a truncated half-normal distribution, and the discrete mass is then smoothed into a continuous function using a Gaussian kernel. 

To model the radial positions of the subhalos, we start with the cumulative distribution function (CDF) of the positive half of a normal distribution (since $r \ge 0$),
$F(r) = \operatorname{erf}\left(\frac{r}{\sqrt{2}\,\sigma}\right), $ where $\sigma$ is the standard deviation that determines the spread of the distribution. To confine the distribution to the desired interval $[r_{\text{inner}},\, r_c]$, we define the truncated CDF as
\begin{equation}
 F_{\text{trunc}}(r) = \frac{\operatorname{erf}\left(\frac{r}{\sqrt{2}\,\sigma}\right) - \operatorname{erf}\left(\frac{r_{\text{inner}}}{\sqrt{2}\,\sigma}\right)}{\operatorname{erf}\left(\frac{r_c}{\sqrt{2}\,\sigma}\right) - \operatorname{erf}\left(\frac{r_{\text{inner}}}{\sqrt{2}\,\sigma}\right)}.   
\end{equation}
To deterministically generate $N$ positions, we select $u_i$ uniformly spaced in the interval $[0,1]$ and then invert the truncated CDF:
\begin{equation}
    \begin{split}
     &r_i = \sqrt{2}\sigma\operatorname{erfinv}(F(\sigma,r_c,r_{\rm inner})),\hspace{0.2cm}\text{where}\\
     &F(\sigma,r_c,r_{\rm inner})=u_i \Bigl[ \operatorname{erf}\!\left(\frac{r_c}{\sqrt{2}\,\sigma}\right) - \operatorname{erf}\!\left(\frac{r_{\text{inner}}}{\sqrt{2}\,\sigma}\right) \Bigr] \\& +\operatorname{erf}\!\left(\frac{r_{\text{inner}}}{\sqrt{2}\,\sigma}\right),\quad\hspace{0.1cm}\text{with}\hspace{0.2cm} i=0,\dots,N-1.   
    \end{split}
\end{equation}
A larger value of $\sigma$ yields a wider distribution, causing the subhalos to be more spread out along the interval. Conversely, a smaller $\sigma$ concentrates the subhalos closer to $r_{\text{inner}}$.

Each subhalo of mass $m$ is spatially distributed using a Gaussian kernel:
\begin{equation}
  K(\vec{r}; h) = \frac{1}{(h\sqrt{2\pi})^3}\exp\!\left(-\frac{|\vec{r}|^2}{2h^2}\right),  
\end{equation}
where $h$ is the smoothing parameter (or \emph{bandwidth}). The resulting continuous density function is given by
\begin{equation}
  \rho(\vec{r}) = \sum_{i=1}^{N} m\, K(\vec{r}-\vec{r}_i; h).  
\end{equation}
The parameter $h$ determines the spatial extent over which each subhalo “spreads” its mass and so it is related with the subhalo radius scale $h\sim R_{\rm SH}/2$. A small value of $h$ indicates that each subhalo's influence is more localized, resulting in a density function with sharp variations. In contrast, a larger $h$ produces greater smoothing, distributing the mass continuously and regularly, meaning that the subhalos are bigger.

The accumulated mass within a sphere of radius $r$ is obtained by integrating the density, and it defines the circular velocity,
\begin{equation}
   M(r) = \int_{|\vec{r}'|\le r} \rho(\vec{r}')\, d^3\vec{r}'\rightarrow v_{\rm SH}(r) = \sqrt{\frac{G\,M(r)}{r}}.
\end{equation}

To correctly define $M(r)$, we have to recover the dimensional values of the boson mass. For a dimensionless scalar boson star solution with a typical field frequency of $w/\mu_{\rm S}\sim 0.87$, which in our setup corresponds to $\bar{\Phi}_c \sim 0.05$ and a dimensionless mass of $M \sim 0.85$ (see Fig.~\ref{masas_BS} in Appendix~\ref{AppendixA}), we target dimensional masses in the range $M_{\rm N} \in [1, 10^6] \,M_{\odot}$. Using the rescaling relations provided in Section~\ref{rescalingsect}, we find that the boson mass must lie in the range $\mu_{\rm S} \in [10^{-22}, 10^{-16}]$~MeV. With this in mind, we explore different subhalo configurations by varying both the boson mass within this range and the number of subhalos, in order to identify the best fit to the observational data.

In this work, we assume a population of equal-mass bosonic subhalos for simplicity and computational capability. While this is an idealization, it enables us to assess the model's viability with a minimal set of free parameters. Future studies will explore more realistic subhalo mass distributions, which may further enhance the model’s ability to reproduce fine features in galactic rotation curves.

%%%%%%%%%%%%%%%%%%%%%%%%%%%%%%%%%%%%%%%%%%%%%%%%%%%%%%%%%%%%%%
\section{\textbf{Results}}
\label{results}
%%%%%%%%%%%%%%%%%%%%%%%%%%%%%%%%%%%%%%%%%%%%%%%%%%%%%%%%%%%%%%

The galaxy sample used in this work consists of 10 objects, previously studied in \cite{Schunck:1998nq}, selected for being both representative of diverse galactic morphologies (spirals and dwarfs) and for their high dark matter content. This makes them particularly suitable for testing alternative dark matter models such as bosonic halos. We adopted the same sample in our previous study \cite{Mourelle:2024dlt}, where we analyzed single-component bosonic models. Using the same set here enables direct comparison with the new two-component framework.

\begin{figure*}[t]
    \centering
    \includegraphics[width=0.4\textwidth]{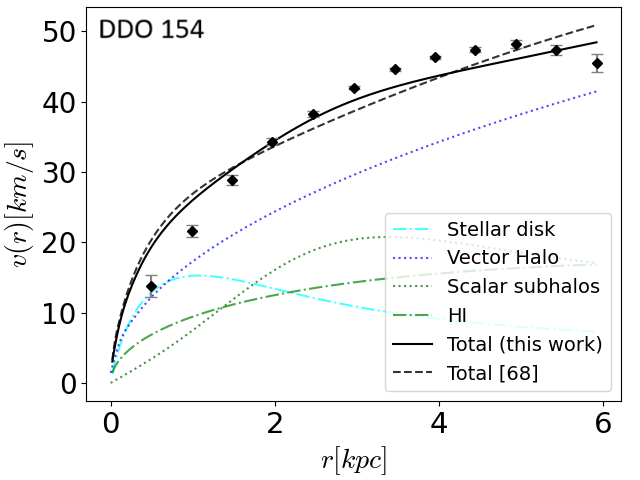}
    \includegraphics[width=0.4\textwidth]{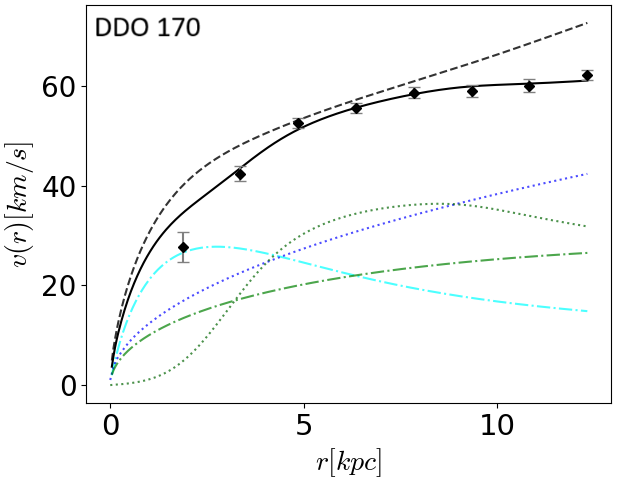}\label{fig:f1}
\\
    \includegraphics[width=0.4\textwidth]{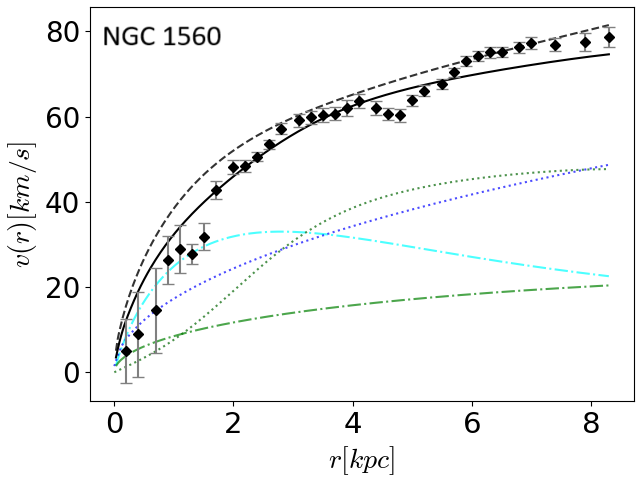}
    \includegraphics[width=0.4\textwidth]{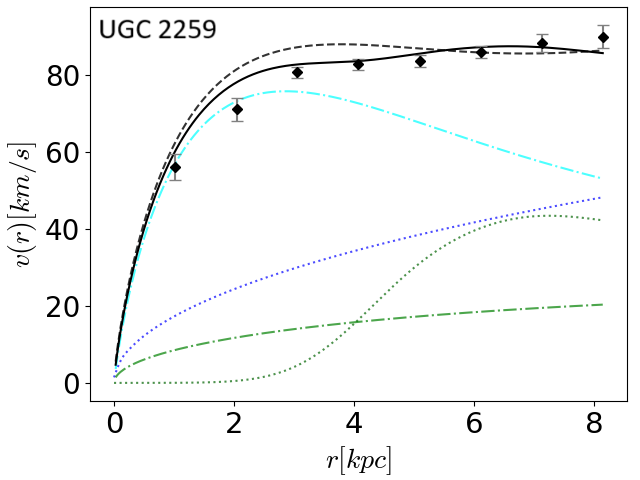}\label{fig:f2}   
    \caption{Rotation curves for the following galaxies: DDO 154 (top left), DDO 170 (top right), NGC 1560 (bottom left), and UGC 2259 (bottom right). Black diamonds indicate observational data for the rotational velocity and are displayed with their respective error bars. Each line indicates the corresponding contribution of a given galactic component to the rotational velocity (see legend) with the black solid line displaying the total contribution obtained with the present model and the black dashed line the corresponding fit from~\cite{Mourelle:2024dlt}. With the new set of observational data, the field frequencies used in the fits of~\cite{Mourelle:2024dlt} have changed to the following values: $w/\mu_{\rm V}=0.9996$ for  DDO 170 and $w/\mu_{\rm V}=0.9997$ for UGC 2259.}
    \label{galaxy1}
\end{figure*}

\begin{figure*}[t]
    \centering
    \includegraphics[width=0.4\textwidth]{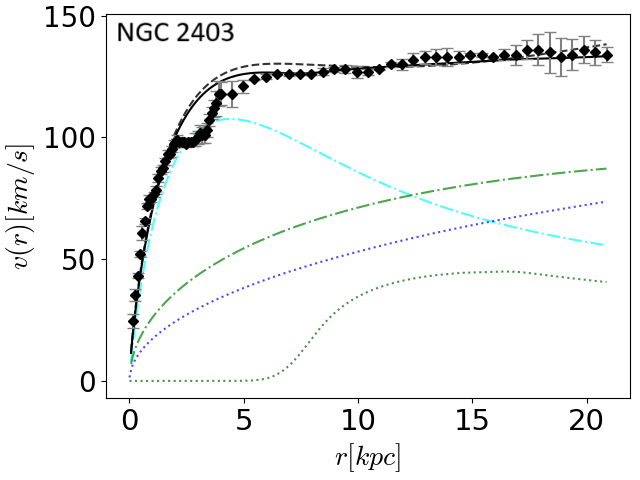}
    \includegraphics[width=0.4\textwidth]{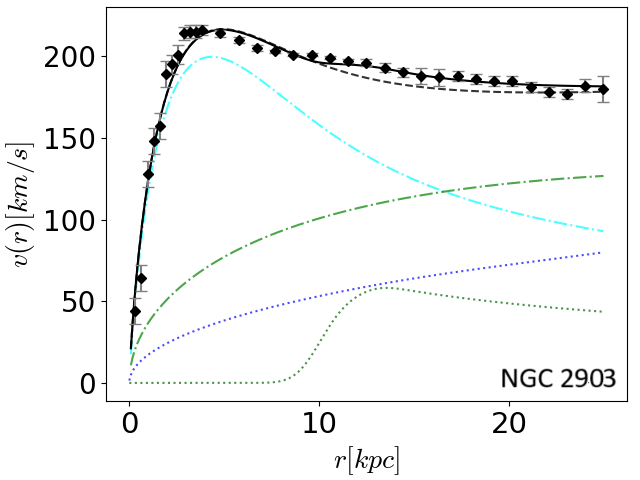}\label{fig:f3}
\\
    \includegraphics[width=0.4\textwidth]{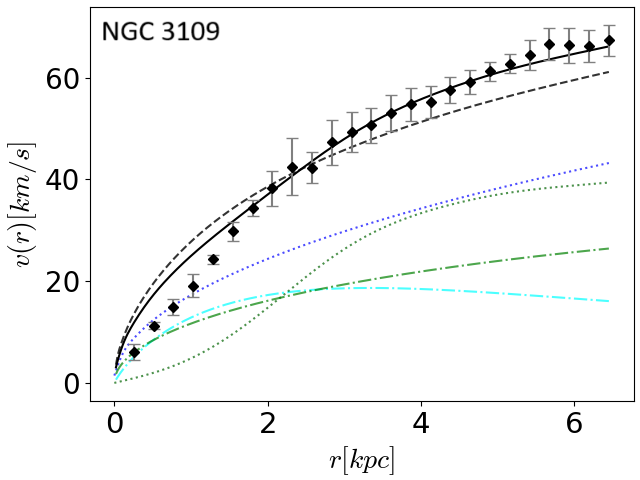}
    \includegraphics[width=0.4\textwidth]{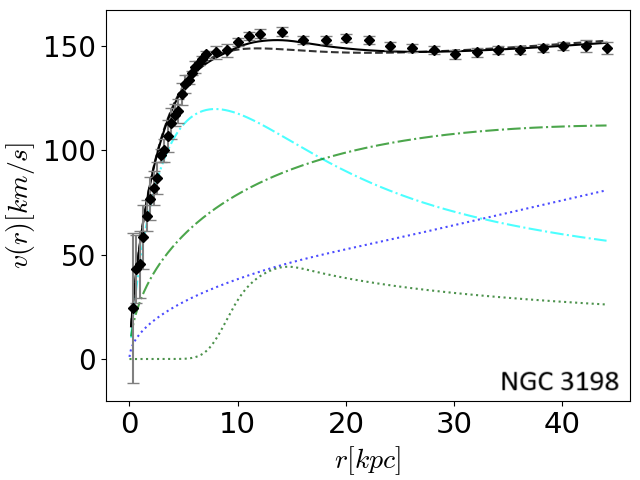}\label{fig:f4}
    \caption{Same as Fig.~\ref{galaxy1} but for galaxies NGC 2403 (top left), NGC 2903 (top right), NGC 3109 (bottom left), and NGC 3198 (bottom right).}
    \label{galaxy2}
\end{figure*}

\begin{figure*}[t]
    \centering
    \includegraphics[width=0.4\textwidth]{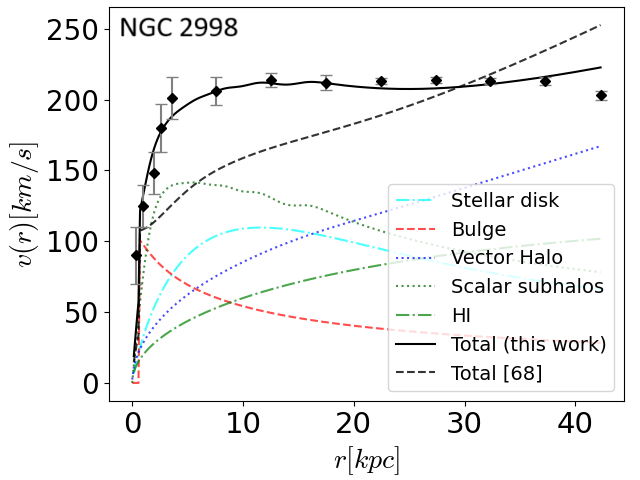}
    \includegraphics[width=0.4\textwidth]{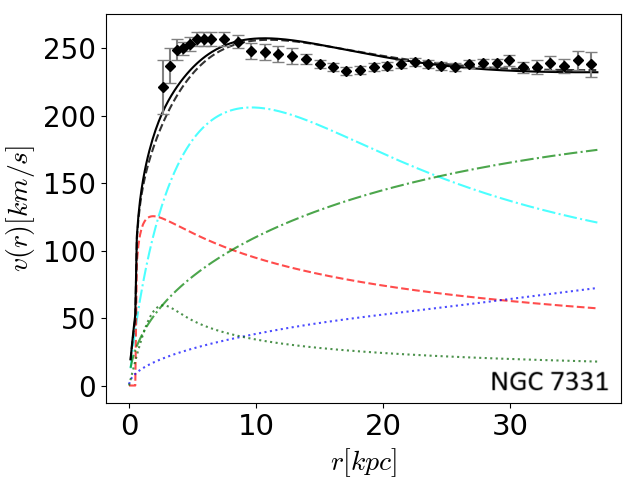}\label{fig:f1}
    \caption{Same as Fig.~\ref{galaxy1} but for galaxies NGC 2998 (left) and NGC 7331 (right). 
    With the new set of observational data, the field frequency for NGC 2998 used in the fit of~\cite{Mourelle:2024dlt} has changed to $w/\mu_{\rm V}=0.9994$.}
    \label{galaxy3}
\end{figure*}

Their names and properties are reported in the first block of entries of Table~\ref{table1_subhalos}. It is important to notice that we use here a more up-to-date set of observational data than in~\cite{Mourelle:2024dlt}, obtained from the repository~\texttt{SPARC}~\cite{sparc2016} and from~\cite{Gentile:2010nb}. While, in general, the data from~\cite{Mourelle:2024dlt} and the new data set agree to good accuracy, for some of the galaxies the observational data is slightly changed. This has some implications in the comparison of the new model with the fits reported in \cite{Mourelle:2024dlt}. In particular, some of the previous values for the frequency of the vector field are now different. The specific instances when this happens are indicated in the captions of Fig.~\ref{galaxy1} and Fig.~\ref{galaxy3}. A further enhancement of the observational data implemented in this work includes the explicit display of the estimated observational errors for the RC, also obtained from~\cite{sparc2016,Gentile:2010nb}. These uncertainties are represented by error bars in Figs.~\ref{galaxy1}-\ref{galaxy3} for all galaxies and they provide valuable insight into how closely our models match the observational results.

%\jcm{In \cref{table1_subhalos}, we display the data from Table I of \cite{Mourelle:2024dlt} with updated values for some magnitudes, and supplemented with the corresponding model properties from our new approach. However, it is important to notice that we have used a more updated set of data for the observational data. Rather than employ the same observed RC values as in \cite{Mourelle:2024dlt}, we have used a more complete and current set of data, obtained from the repository \textit{SPARC} \cite{sparc2016} and further works \cite{Gentile:2010nb}. In general, our former data and the new data set coincide extremely well, but some of the studied galaxies present some changes in the observational data. This will have some implications when comparing the new model approach with the one in \cite{Mourelle:2024dlt}. As the real data suffer small changes, some values for the vectorial field frequency, using the approach in\cite{Mourelle:2024dlt} have to change now. We will indicate the cases when this happens, as we will compare our new models with the previous paper's ones.}

%\jcm{Further enhancements related to observational data have been implemented in this work, as we now explicitly show the estimated observational errors. These uncertainties, represented by error bars, have been included for all galaxies since they provide valuable insight into how closely our models match reality. Again, these error values are obtained from the same sources as the observational data, namely \cite{sparc2016,Gentile:2010nb}.}

We use the $L_2$ norm to quantify whether our new model yields an improved description of the observed galactic RC compared to the fits obtained in~\cite{Mourelle:2024dlt}. The $L_2$ norm, commonly known as the Euclidean norm, is an absolute statistical measure that quantifies the magnitude of a vector $\mathbf{e}$ as,
\begin{eqnarray}
    \|\mathbf{e}\|_2 = \sqrt{\sum_{i=1}^{n} e_i^2}\,.
\end{eqnarray}
When comparing theoretical galactic RC to the observed ones, the $L_2$ norm serves as a measure of the goodness of the model. Specifically, if the observed rotation velocities at various radii across the galaxy are represented by $v_i^{\text{obs}}$ and the corresponding velocities predicted by our model are $v_i^{\text{mod}}$, then the deviation between model and data can be quantified as
\begin{eqnarray}
   L_2= \|\mathbf{e}\|_2 = \sqrt{\sum_{i=1}^{n}\left(v_i^{\text{obs}} - v_i^{\text{mod}}\right)^2}\,.
\end{eqnarray}
Minimizing the $L_2$ norm provides a statistical criterion for assessing and comparing various models, facilitating the identification of the model most consistent with the observational data.

%In our case vector $\mathbf{e}$ is a vector of residuals between the observational data and our theoretical model, $\mathbf{e} = (e_1, e_2, \dots, e_n)$.

%\jcm{Since our goal is to assess whether our new model yields a more consistent description of the observed galactic RC compared than that in~\cite{Mourelle:2024dlt}, we quantify the improvement using an absolute statistical measure: the $L_2$ norm.}

%\jcm{The L$_2$ norm, commonly known as the Euclidean norm, quantifies the magnitude of a vector. Given a vector of residuals between observational data and a theoretical model, $\mathbf{e} = (e_1, e_2, \dots, e_n)$, the L$_2$ norm is mathematically defined as:
%\begin{equation}
%    \|\mathbf{e}\|_2 = \sqrt{\sum_{i=1}^{n} e_i^2}.
%\end{equation}}

%\jcm{In our context, particularly when comparing theoretical rotation curves of galaxies to the observed rotation velocities, the L$_2$ norm serves as a measure of the goodness of a model. Specifically, if the observational velocities at various radii are represented by $v_i^{\text{obs}}$, and the corresponding velocities predicted by a model are $v_i^{\text{mod}}$, then the deviation between model and data can be quantified as:
%\begin{equation}
%   L_2= \|\mathbf{e}\|_2 = \sqrt{\sum_{i=1}^{n}\left(v_i^{\text{obs}} - v_i^{\text{mod}}\right)^2}.
%\end{equation}
%Minimizing this value provides a statistical criterion for assessing and comparing various models, facilitating the identification of the model most consistent with observational data.
%}

As stated above, a key difference with~\cite{Mourelle:2024dlt} is that we now fix the vector boson mass $\mu_{\rm V}=9\times 10^{-26}\,\mathrm{eV}$ across all galaxies, varying only the ratio $w/\mu_{\rm V}$ per model. For the distribution of the scalar bosonic subhalos, we find a degeneracy in our results (i.e.~equally admissible models) between the number of subhalos $N$ and the mass of each structure $M_N$. Therefore, we choose to fix the product $N\times M_N$ to a constant value and study the viability of the corresponding model by fixing the mass of the boson, which sets the size and mass scales for the subhalos, and taking a value for $N$ at the same time. We statistically determine the optimal value of $N \times M_N$ as the one that  minimizes the corresponding $L_2$ norm. These values are reported in Table~\ref{table1_subhalos} along with the $L_2$ norms for the current model and for the one used in~\cite{Mourelle:2024dlt} (the latter indicated in the last two columns by $L_2$-n and $L_2$-o, respectively). The lower the $L_2$ norm, the better the model. As shown in Table~\ref{table1_subhalos} the comparison of the $L_2$ norms for all galaxies of our sample indicates that the new model constitutes a significant improvement with respect to~\cite{Mourelle:2024dlt}, as all $L_2$-n values are lower than the corresponding $L_2$-o ones. In some cases this reduction is particularly large as in e.g.~NGC 1560, NGC 2998, and NGC 7331. This shows that our new approach based on two-component halos (vector DM for the main one and scalar DM for the subhalos) enhances the agreement with observation compared to the model of~\cite{Mourelle:2024dlt}.

A complementary metric to assess our model is the absolute percentage error, which provides a measure of the deviation of the estimated value from the real one,
\begin{equation}
    \varepsilon_i^{\mathrm{perc}} = 100 \cdot \left| \frac{v_{\rm T}(r_i^{*}) - v_i^{\mathrm{real}}}{v_{\rm T}(r_i^{*})} \right|\,.
    \label{errorsperc}
\end{equation}
This error metric is useful when evaluating the magnitude of the deviation in relation to the scale of the real data. Using this estimator, we can identify the points in the model with the largest and smallest deviations from the observational data, denoted by $E_{\text{max}}$ and $E_{\text{min}}$, respectively. Additionally, by summing all point-wise errors and dividing by the total number of points, we obtain the average error for each model $\Delta E=n^{-1}\sum_i^n \varepsilon_i^{\text{perc}}$, gaining further insight into the model’s overall performance. The values of these three quantities for all galaxies are also reported in Table~\ref{table1_subhalos}. 

Our two-component bosonic dark matter model maintains a balance between physical motivation and fitting efficiency. The main halo is modeled with a fixed vector boson mass $\mu_V$ and a nearly constant field frequency $\omega$, consistent with the sizes and velocities of the observational curves. Scalar bosonic subhalos are described via a phenomenological model, where the total mass $N \times M$ and a few structural parameters (radial extent, spread, and core size) are, again, constrained by the measured rotation curves and soliton scaling relations~\cite{Schive:2014dra, Marsh:2015xka}. Parameter variations across galaxies reflect physical differences rather than overfitting, and all parameters are tied to observables or theoretical priors, ensuring model consistency.

Despite some galaxies attain high values of $E_{\rm max}$ (e.g.~NGC 1560 and NGC 2998), the average error $\Delta E$ mostly remains moderate and  comparable to the observational errors, ranging between $2-10\,\mathrm{km/s}$.

\begin{table*}
\caption{The first block of columns reports the parameters of our galaxy sample, including galaxy name (and reference), luminosity $L$, effective radius $R_{\rm disk}$, mass disk/luminosity ratio $M_{\rm disk}/L$,  HI mass $M_{\rm HI}$, mass bulge/luminosity ratio $M_{\rm bulge}/L$, and HI/He ratio. The numerical values for these quantities are derived from studies of mass distributions and the photometric and dynamical properties of spiral and dwarf galaxies \cite{Ashman1988, Blanton2013, Read2003, Trentham2003, Sofue1997, Elmegreen2016, Silk2002, 1996A&AS..118..557D,Kormendy2013,Schunck:1998nq}. 
The second block reports the values of the model parameters, namely the 
product of the number of scalar subhalos and the mass of each subhalo, $N\times M_N$, spatial spread and size of the subhalos, $h$ and $\sigma$, characteristic radii where subhalos are located, $r_{\rm in}$ and $r_{\rm ext}$, and field frequency for the vector boson, $w_M$. The third block of columns shows the statistical estimators, namely the minimum, maximum, and average percentual errors for each galaxy, $E_{\rm min}$, $E_{\rm max}$, and $\Delta E$, and the $L_2$ norms for both the model of the present work, $L_2$-n, and that from \cite{Mourelle:2024dlt}, $L_2$-o.  }
\begin{tabular}{c|cccccc|cccccc|ccccc}
\hline
\textbf{Galaxy}& $L$ & $R_{\rm disk}$ & $\frac{M_{\rm disk}}{L}$ & $M_{\rm HI}$ & $\frac{M_{\rm bulge}}{L}$ &$ \frac{\rm HI}{\rm He}$ & $N\times M_N$ & $h$ & $ \sigma$ & $r_{\rm in}$ & $r_{\rm ext}$ & $w_M$ & $E_{\rm min}$& $E_{\rm max}$ & $\Delta E$ & $L_2$-n& $L_2$-o\\
 & $(\mathrm{10^9\,M_{\odot}})$ 
 & (kpc) 
 & 
 & $(\mathrm{10^6\,M_{\odot}})$ 
 & 
 & 
 & 
 &
 &
 &
 &
 & $(10^{-3})$
 &
 &
 &
 &
 &
\\
\hline

DDO 154 \cite{1988ApJ...332L..33C} & $0.05$ & $0.50$ & $1.4$ & $5.0$ & $0$ & $1.5$  &$4\times 10^{8}$ & $0.8$ & $1$ & $1.7$ & $3$ & $999.7$ & $0.34$ & $28.81$ & $7.12$ & $1.59$ &  $4.76$\\
DDO 170 \cite{1990AJ.....99..547L} & $0.16$ & $1.28$ & $3.7$ & $6.0$ & $0$ & $1.3$ & $29\times 10^{8}$ & $0.8$ & $8$ & $3$ & $9$ & $999.8$ & $0.03$ &$19.41$ & $3.58$ &$6.36$ & $46.81$\\
NGC 1560 \cite{Kormendy2013} & $0.35$ & $1.30$ & $3.0$ & $3.5$ & $0$ & $1.6$ & $50\times 10^{8}$ & $0.9$ & $14$& $2$ & $9$ &  $999.7$& $0.11$ & $58.63$& $9.63$& $20.90$ & $174.69$  \\
UGC 2259 \cite{santos2020baryonic}  & $1.02$ & $1.33$ & $4.5$ & $4.0$ & $0$ & $1.4$& $33\times 10^{8}$ & $0.7$ & $6$& $4$ & $7$ &  $999.7$ & $1.52$ & $9.01$ & $3.86$& $12.26$ & $24.95$\\
NGC 2403 \cite{dors2024cosmic} & $7.90$ & $2.05$ & $2.00$ & $43.0$ & $0$ & $1.2$& $80\times 10^{8}$ & $0.7$ & $10$& $8$ & $16$ &   $999.7$& $0.09$ & $31.26$ & $6.17$ & $111.05$ & $230.36$\\
NGC 2903 \cite{Silk2002} & $15.3$ & $2.02$ & $3.50$ & $61.0$ & $0$ & $1.7$& $11\times 10^{9}$ & $0.7$ & $6$& $10$ & $13$ &  $999.7$& $0.09$ & $33.81$& $3.84$ & $14.58$ & $97.50$\\
NGC 2998 \cite{chattopadhyay2006objective}  & $12.0$ & $5.40$ &$ 0.25$ & $20.0$ & $1.5$ & $1.5$& $60\times 10^{9}$ & $1.0$ & $10$& $1$ & $15$ &  $999.5$& $0.03$ & $176.22$& $16.79$ & $36.92$& $344.71$ \\
NGC 3109 \cite{garling2024dual}& $0.81$ & $1.55$ & $0.4$ & $7.5$ & $0$ & $1.4$ & $4\times 10^{9}$ & $0.9$ & $10$& $2.8$ & $9$ &  $999.7$& $0.11$ & $48.76$ & $7.88$ & $21.92$ & $26.14$\\
NGC 3198 \cite{Reste:2023zpq} & $9.00$ & $3.68$ & $3.83$ & $37.0$ & $0$ & $1.6$ & $7\times 10^{9}$ & $1.0$ & $10$& $9$ & $14$ &  $999.8$ & $0.12$& $21.03$ & $3.83$ & $52.90$ & $59.37$ \\
NGC 7331 \cite{Smith2021} & $54.0$ & $4.48$ & $2.25$ & $80.0$ & $1.7$ & $1.3$& $55\times 10^{8}$ & $0.9$ & $6$ & $0.8$ & $2$ &  $999.8$& $0.37$& $9.45$&$3.54$ &$7.49$ &$108.24$ \\
\hline
\end{tabular}
\label{table1_subhalos}
\end{table*}

%\end{widetext}

Finally, Figs.~\ref{galaxy1}-\ref{galaxy3} display the RC of the selected observational sample. More precisely, Fig.~\ref{galaxy1} shows the curves for galaxies DDO 154, DDO 170, NGC 1560, UGC 2259 and Fig.~\ref{galaxy2} for galaxies NGC 2403, NGC 2903, NGC 3109, NGC 3198. These two sets of galaxies share the common feature of lacking a bulge component. In contrast, galaxies NGC 2998 and NGC 7331, shown in Fig.~\ref{galaxy3}, exhibit a significant bulge contribution to their luminous matter content. In each plot we display, as black dashed lines, the RC fits of~\cite{Mourelle:2024dlt} using only the vectorial DM components. While these fits are for the most part identical to those presented in our previous paper, for some of the galaxies we have updated the field frequency associated with the vector boson in order to use improved results from observational data (see figure captions for details). In agreement with the quantitative results reported in Table~\ref{table1_subhalos}, Figs.~\ref{galaxy1}-\ref{galaxy3} show that the new model (black solid lines) yields improved adjustments across all galaxies compared to the model from~\cite{Mourelle:2024dlt}. In addition, most of the new RC are within the error bars of the observational data, and in some cases remarkably so (e.g.~NGC 2409 or NGC 3198).

%\jcm{Due to qualitative comparison reasons, in each plot we show, as black dashed lines, the models obtained using only the vectorial DM component, as in \cite{Mourelle:2024dlt}. Again, we emphasize that although these dashed models are generally identical to those presented in the cited paper, for certain galaxies, the field frequency associated with the vectorial DM model has changed due to updates in observational data. We specify the updated field frequencies for these previously published models only when the original values from \cite{Mourelle:2024dlt} required adjustment. }

%%%%%%%%%%%%%%%%%%%%%%%%%%%%%%%%%%%%%%%%%%%%%%%%%%%%%%%%%%%%%%
\section{\textbf{Discussion}}
\label{conclusions}
%%%%%%%%%%%%%%%%%%%%%%%%%%%%%%%%%%%%%%%%%%%%%%%%%%%%%%%%%%%%%%

Building on~\cite{Mourelle:2024dlt} we have explored the possibility of using vector and scalar bosons to model the components of galactic DM halos and subhalos in order to reproduce the observed rotation curves of galaxies. The main halo of this two-component DM model is generated by a a spinning, axisymmetric vector boson field configuration in the non-relativistic limit $w/\mu_{\rm V} \rightarrow 1$. The secondary halo component consists of a population of static scalar boson ``stars'' representing intermediate galactic structures. Within this framework, we have set the vector boson mass to $\mu_{\rm V} = 9\times 10^{-26}\,\mathrm{eV}$, while allowing the scalar boson mass to vary in the range $\mu_{\rm S} \in [10^{-16}, 10^{-22}]\,\mathrm{MeV}$.
We have further refined our previous analysis~\cite{Mourelle:2024dlt} by incorporating updated observational data from the \texttt{SPARC} repository. This has allowed us to explicitly include observational error bars for all galaxies considered and assess the goodness of the model in a quantitative way. Employing a statistical evaluation based on the $L_2$ norms and percentage errors between our models and the observational data, we have defined a clear criterion to determine the best set of parameters for each model (i.e.~the optimal number of scalar subhalos and the mass of each subhalo), as the one minimizing these statistical measures. By comparing our new model with the one used in~\cite{Mourelle:2024dlt}, we have observed that our two-component DM approach notably improves the agreement between observations and theoretical predictions. Moreover, the ad-hoc fits for the galactic RC reported in our previous work are now justified on physical grounds. 

Fixing the mass of the vector boson has addressed an issue found in~\cite{Mourelle:2024dlt} where it had to be adapted for each galaxy. While in our previous work an estimated value of $\mu_{\rm V}=9\times 10^{-26}\,\mathrm{eV}$ was obtained by averaging the different values used in the galaxy sample, using this single value sets a constraint for the mass of the vector boson that might play a cosmological role. This value can be interpreted as the minimum mass of the vector boson needed to model galactic DM halos as a vector field coupled to gravity. With this fully constrained value, our model only requires the addition of a distribution of subhalos to complete the total DM of a given galaxy. These subhalos play a secondary role but are crucial to precisely adjust our models to the observed RC. The degeneracy we have found between the number of subhalos and the mass of each of them has not allowed us to obtain a single value for the mass of the scalar bosonic particle. We have seen that what determines the accuracy of the model is the product $N\times M_N$, thus having the possibility of varying either quantity. Using values for the masses of the scalar subhalos similar to those derived in the case of primordial black holes (PBHs), assuming those can range from a few solar masses to $10^6\,M_{\odot}$, has allowed us to constrain the mass of the scalar boson in the range $10^{-22}\,\mathrm{MeV} \leq \mu_{\rm S} \leq 10^{-16}\,\mathrm{MeV}$.

Our model for DM subhalos based on boson stars can be confronted with other alternatives, in particular, PBHs. Boson stars share with PBH their compact, non-baryonic nature and the fact that they are non-luminous and challenging to detect directly. This makes some observational constraints developed for PBHs potentially applicable to boson stars as well. Since both affect galactic dynamics solely through gravity, they can similarly influence RC, gravitational lensing, and structure formation. If one assumed an extended mass distribution for boson stars (e.g.~Gaussian or log-normal centered around $10^2$ to $10^3~M_\odot$) then the analysis of the observational constraints could closely follow the framework used for PBHs with extended mass functions~\cite{Carr2020}. However, our model is less elaborate, since we are assuming a monochromatic mass distribution with a degeneracy in the mass and number of objects.  Moreover, the applicability of such constraints depends on the specific properties of the boson stars. For instance, if their masses and densities were within the sensitivity range of microlensing surveys like EROS, OGLE, or MACHO, then the same microlensing bounds on PBHs~\cite{Torres2000} could apply. If boson stars were more diffuse, they could potentially evade these bounds. In our case, boson stars can have masses ranging from a few solar masses to millions of them, which makes our model prone to possibly evade the PBH constraints in the range $M>10^3$~$M_{\odot}$. Unlike PBHs, purely gravitational boson stars without significant accretion may not affect the cosmic microwave background (CMB), thereby avoiding those specific constraints as well~\cite{Carr2020}. Their distribution within galactic halos must also be considered. Depending on number density, individual mass, and average separation, boson stars might disturb the galactic halo substructure, which can be tested using N-body simulations or kinematic data~\cite{Schunck2003}. Additionally, mergers of boson stars, although typically produce weaker gravitational wave signals than PBH mergers, could be constrained if the population is sufficiently massive and merger rates are high~\cite{Palenzuela2007}. Finally, our model assumes that the subhalo structures are a complementary component to the total DM halo (a smaller contribution to the principal vector DM halo). This also allows us to evade some MACHO constraints given by microlensing and further observational tools in the range of masses $1-10^3 M_{\odot}$, as we are not regarding the subhalos as the main DM contribution but as a second component \cite{Brandt:2016aco,Carr:2016drx,Espinosa:2017sgp,Carr2020}.

The nature of DM remains a widely debated topic. Many contemporary models, including those based on baryonic or non-relativistic bosonic matter, aim to address the galactic RC problem. Within our framework, we have proposed a viable solution using a unified, gravitationally coupled theory of two bosonic fields. We believe this model offers valuable insights into fundamental physics and may have potential cosmological implications. An extension of this model we plan to undertake will incorporate angular momentum into the subhalo distributions along with variety of bosonic solutions, rather than a single configuration. The results from those studies will be reported elsewhere.

%\jcm{END}
%A promising extension of this model would be to introduce angular momentum into the subhalo distribution and to consider a variety of bosonic solutions, rather than a single repeated configuration. The nature of dark matter remains a widely debated topic. Many contemporary models, including those based on baryonic or non-relativistic bosonic matter, aim to address the RC problem. Within our framework, we propose a viable solution using a unified, gravitationally coupled theory of two bosonic fields from the same family. We believe this model offers valuable insights into fundamental physics and the cosmological implications of General Relativity.

%%%%%%%%%%%%%%%%%%%%%%%%%%%%%%%
\begin{acknowledgements}
JCM thanks J.J.~Blanco Pillado for useful comments and helpful discussions. JCM 
%is supported by the project ``Dark matter exotic compact objects:~a gravitational-wave catalog of bosonic star mergers in the LIGO-Virgo-KAGRA era.” (Ref.~CPI-25-030) from the Spanish Ministry of Science and Innovation via the Ram\'on y Cajal programme (grant RYC2022-037424-I), funded by MCIN/AEI/10.13039/501100011033 and by ``ESF Investing in your future”. 
and NSG acknowledge support from the Spanish Ministry of Science and Innovation via the Ram\'on y Cajal programme (grant RYC2022-037424-I), funded by MCIN/AEI/10.13039/501100011033 and by ``ESF Investing in your future”. NSG and JAF are supported by the Spanish Agencia Estatal de Investigaci\'on (grant PID2021-125485NB-C21) funded by MCIN/AEI/10.13039/501100011033 and ERDF A way of making Europe, and by the Generalitat Valenciana (grant CIPROM/2022/49). We acknowledge further support from the European Horizon Europe staff exchange (SE) programme HORIZON-MSCA2021-SE-01 Grant No. NewFunFiCO-101086251.
\end{acknowledgements}
%%%%%%%%%%%%%%%%%%%%%%%%%%%%%%%%

%%%%%%%%%%%%%%%%%%%%%%%%%%%%%%%%%%%%%%%%%%%%%%%%%%%%
\appendix
\section{ Equilibrium equations for the EKG and EP systems}
\label{AppendixA}
%%%%%%%%%%%%%%%%%%%%%%%%%%%%%%%%%%%%%%%%%%%%%%%%%%%%%

The equilibrium equations for static boson stars (EKG system) are given by
\begin{equation}
 \begin{split}
        & \Phi''=-\left(\frac{2}{r}+\frac{\gamma'-\lambda'}{2}\right)\Phi' + e^{\lambda}\Phi\left(\frac{d V(|\Phi|^2)}{d|\Phi|^{2}}-e^{-\gamma}w^{2}\right),\\
        &  \gamma'\left(r\right) = \frac{e^{\lambda}-1}{r} + 8\pi r\left[\phi'^{2} + e^{\lambda}\left(w^{2}e^{-\gamma}\phi^{2} -V(|\Phi|^2)\right)\right], \\
        &\lambda'\left(r\right) = \frac{1-e^{\lambda}}{r} + 8\pi r\left[\phi'^{2}+e^{\lambda}\left(w^{2}e^{-\gamma}\phi^{2} +V(|\Phi|^2)\right)\right]. 
    \end{split}
    \label{equilibriumBS}
\end{equation}
These equations require boundary conditions ensuring regularity at the origin and asymptotic flatness. The resulting eigenvalue problem is solved via a one-parameter shooting method, using rescaled parameters:
\begin{equation}
\bar{\Phi}_0=\sqrt{4\pi}\Phi_0,\quad \bar{r}=\mu_{\rm S} r,\quad \bar{t}=\mu_{\rm S}t\,.
\end{equation}
Boson stars lack a sharp physical boundary, with their density smoothly decreasing at large radii. Practically, the star boundary is approximated where $99\%$ of its mass is contained, although no strict surface exists. Employing a Schwarzschild-like metric allows us to relate the metric function to the ADM mass $M$, and thus we define the mass function:
\begin{equation}
M(r)=\frac{r}{2}\left(1-\frac{1}{e^{\lambda(r)}}\right),
\label{mass_St_BS}
\end{equation}

\begin{figure*}[t!]
    \centering
    \includegraphics[width=0.42\textwidth]{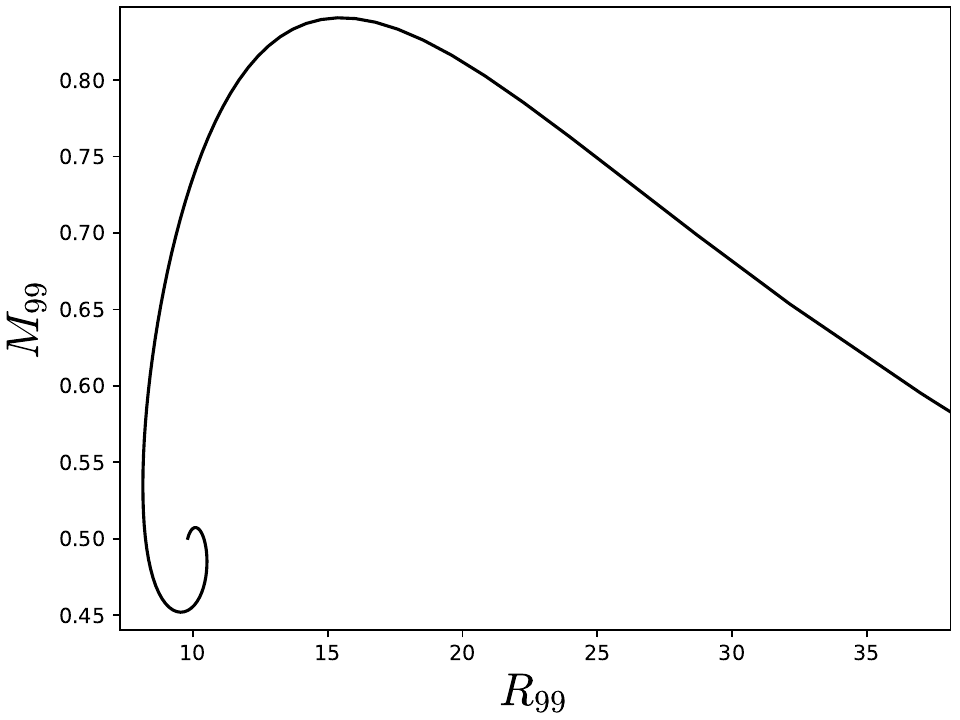}
    \includegraphics[width=0.42\textwidth]{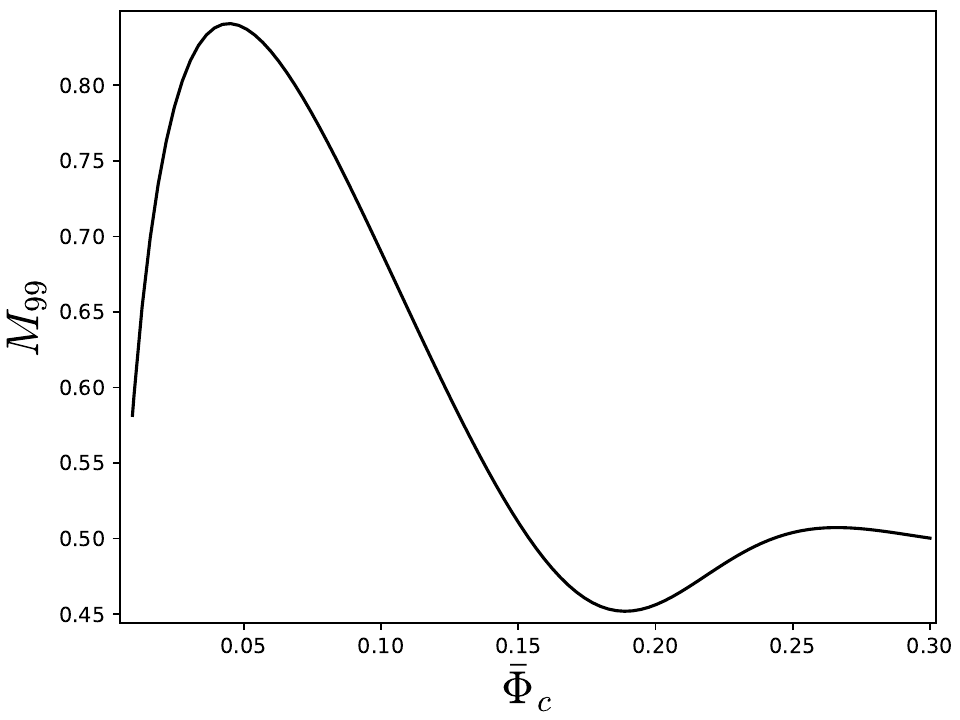}\label{fig:f1}
    \caption{Left panel: mass-radius relation for static boson stars. Expressed in dimensionless units scaled by the boson mass, the corresponding physical masses and radii can be recovered using the relations provided in \cref{rescalingsect}. We clearly distinguish between the primary solution branch, located in the upper region, and the secondary branch—often associated with instabilities—situated in the lower region. Right panel: mass as a function of the field amplitude at the center. This plot illustrates how the relationship between the central field value and the solution is unique. }
    \label{masas_BS}
\end{figure*}

\vspace{0.5cm}

which quantifies mass distribution within radius $r$. This framework enables analyzing the behavior of mass, field, and metric functions with respect to the radial coordinate. Fig.~\ref{masas_BS} shows the mass-radius and mass-central field curves for our boson star solutions used in the DM subhalos.

In regards to the EP system we note that, for numerical reasons, the field equations have to be written in a specific form. To this end, we employ both the standard Einstein field formalism and the Euler–Lagrange approach. The equations for the metric functions are derived within the framework of General Relativity. However, to obtain the equations governing the Proca field, instead of relying on the Einstein field formulation, we define an effective Lagrangian as follows:
\begin{equation}
 \mathcal{L}_{\rm eff}^{\rm P}  =\sqrt{-g}\left(-\frac{1}{4}\textit{F}_{\alpha\beta}\textit{F}^{*\alpha\beta}-\frac{1}{2}\mu^2\textit{A}_{\alpha}\textit{A}^{*\alpha}+\frac{\mathcal{L}^2}{2}\right) ,
\end{equation}
And the Euler-Lagrange equations are obtained as usual:
\begin{equation}
    \begin{split}
        &V^{\rm EL}=\frac{\partial}{\partial r}\left(\frac{\partial \mathcal{L}_{\rm eff}^P }{\partial V_{,r} }\right)+\frac{\partial}{\partial \theta}\left(\frac{\partial \mathcal{L}_{\rm eff}^P }{\partial V_{,\theta} }\right)-\frac{\partial \mathcal{L}_{\rm eff}^P }{\partial V}=0,\\
        &H_i^{\rm EL}=\frac{\partial}{\partial r}\left(\frac{\partial \mathcal{L}_{\rm eff}^P }{\partial H_{i,r} }\right)+\frac{\partial}{\partial \theta}\left(\frac{\partial \mathcal{L}_{\rm eff}^P }{\partial H_{i,\theta} }\right)-\frac{\partial \mathcal{L}_{\rm eff}^P }{\partial H_i}=0,
    \end{split}
\end{equation}
where $i=1,2,3.$
The combinations of equations that allow us to have the EP system described properly to be used in the solver, is the following:
\begin{equation}
\begin{split}
&e^{2\alpha}\frac{r^2}{2}\sin^2(\theta)\left(E_{t}^{t}-E_{r}^{r}-E_{\theta}^{\theta}+E_{\psi}^{\psi}\right)=0,\\
&e^{2\alpha}\frac{r^2}{2}\sin^2(\theta)\left(E_{t}^{t}+E_{r}^{r}+E_{\theta}^{\theta}-E_{\psi}^{\psi}+\frac{2WE_{\psi}^{t}}{r}\right)=0,\\
&e^{2\alpha}\frac{r^2}{2}\sin^2(\theta)\left(-E_{t}^{t}+E_{r}^{r}+E_{\theta}^{\theta}-E_{\psi}^{\psi}-\frac{2WE_{\psi}^{t}}{r}\right)=0,\\
&2re^{2\nu +2\alpha-2\beta}E_{\psi}^{t}=0,\\
& e^{\beta-\nu}r\sin\theta\left(W\sin\theta \, V^{\rm EL}-H_3^{\rm EL}\right)=0,\\
&e^{-\beta+\nu}\sin\theta\left[V^{\rm EL}+e^{-2\nu+2\beta}\sin\theta W\left(H_3^{\rm EL}\right.\right.\\&
\left.\left.-W\sin\theta V^{\rm EL}\right)\right]=0,\\
&-e^{\nu+2\alpha-\beta}r^2\sin\theta\left(H_2^{\rm EL}+e^{\nu+\beta}H_2\sin\theta E^r_r\right)=0,\\
&-e^{\nu+2\alpha-\beta}r^2\sin\theta\left(H_1^{\rm EL}+e^{\nu+\beta}H_1\sin\theta E^{\theta}_{\theta}\right)=0.
\end{split}
\label{EP-system}
\end{equation}

\vspace{5cm}

%%%%%%%%%%%%%%%%%%%%%%%%%%%%%%%%%%%%%%%%%%%%%%%%%%%%
\section{Approach with the $\chi^2$}
\label{AppendixB}
To determine the most suitable parameters for our approach, we employed a statistical methodology based on minimizing the $L_2$ norm. As discussed in the main text, the corresponding estimator is given by:

\begin{eqnarray}
   L_2= \sqrt{\sum_{i=1}^{n}\left(v_i^{\text{obs}} - v_i^{\text{mod}}\right)^2}\,.
\end{eqnarray}
Minimizing the $L_2$ norm offers a robust statistical criterion for evaluating and comparing different models, enabling the identification of the one that best fits the observational data. However, this methodology becomes less appropriate when each data point is associated with a known observational error. While it can still be useful for comparative purposes in the absence of detailed knowledge about these statistical uncertainties, it is generally preferable to use it only when such information is lacking. When the observational errors are well understood and vary across data points, the $L_2$ estimator is less suitable than the $\chi^2$ statistic. Let us now see how the latter is derived:
\begin{eqnarray}
   \chi^2= \sqrt{\sum_{i=1}^{n}\frac{\left(v_i^{\text{obs}} - v_i^{\text{mod}}\right)^2}{\sigma_i^2}}\,
\end{eqnarray}

where $\sigma_i$ is the observational error associated with each measurement.
In our case, although observational error estimates were available, they originated from multiple sources, and the details of their derivation are not fully known. For this reason, while we present them for comparison, we have chosen to rely primarily on the $L_2$ norm as our estimator. Nonetheless, as a complementary analysis, we now include a study of our models based on minimizing the $\chi^2$ statistic. This allows us to examine how our previous fits are affected, in what ways they differ, and whether the use of either $L_2$ or $\chi^2$ indiscriminately is justified in this context.

In \cref{table1_subhalos_2}, we present the statistical estimators obtained through both methods for comparison purposes. We also show the reduced $\chi^2$ as $\chi^2_{red}$ which is obtained through,
\begin{equation}
   \chi^2_{red}=\frac{\chi^2}{n-k} 
\end{equation}
where $n$ is the number of data points and $k$ is the number of fitted parameters.
This value allows us to have an idea of the goodness  of our model fitting, just by comparing it with the reference:

\begin{align}
    \chi^2_{\text{red}} &\simeq 1 &&\text{Good fit.} \nonumber\\
    \chi^2_{\text{red}} &\gg 1    &&\text{Poor fit, errors underestimated.}\nonumber \\
    \chi^2_{\text{red}} &\ll 1    &&\text{Overfitting, uncertainties overestimated.}\nonumber
\end{align}

\begin{table*}
\caption{ In this table, we present, for the set of galaxies analyzed in this work, the values of the mass per number of subhalos obtained by minimizing the $L_2$ norm, along with the associated estimators introduced previously in \cref{table1_subhalos}. Separated by a double vertical bar, we also include the corresponding values of $N \times M_N$ obtained through $\chi^2$ minimization, together with the associated minimum and maximum errors. Additionally, we report the total $\chi^2$ value as well as the reduced chi-squared, $\chi^2_{red}$.
 }
\begin{tabular}{c|c|ccccc||c|ccccc}
\hline
\textbf{Galaxy}&$(N\times M_N)L_2$ & $E_{\rm min}$& $E_{\rm max}$ & $\Delta E$ & $L_2$-n& $L_2$-o&$(N\times M_N)\chi2$ &$E_{\rm min}$& $E_{\rm max}$ & $\Delta E$ & $\chi^2$ & $\chi^2_{red}$
\\
\hline

DDO 154 \cite{1988ApJ...332L..33C}  &$4\times 10^{8}$ &  $0.34$ & $28.81$ & $7.12$ & $1.59$ &  $4.76$ & $6\times 10^{8}$ & $0.32$ & $29.07$ & $6.83$ & $11.83$ & $1.97$\\
DDO 170 \cite{1990AJ.....99..547L}  & $29\times 10^{8}$  & $0.03$ &$19.41$ & $3.58$ &$6.36$ & $46.81$ &$30\times 10^{8}$ & $0.18$ & $19.42$ & $3.67$ & $2.95$ & $1.47$ \\
NGC 1560 \cite{Kormendy2013} & $50\times 10^{8}$ & $0.11$ & $58.63$& $9.63$& $20.90$ & $174.69$ & $56\times 10^{8}$ & $0.31$ & $58.65$ & $9.17$ & $11.68$ & $0.40$ \\
UGC 2259 \cite{santos2020baryonic}  & $33\times 10^{8}$  & $1.52$ & $9.01$ & $3.86$& $12.26$ & $24.95$ & $31\times 10^{8}$ & $0.87$ & $9.00$ & $3.89$ & $3.93$ & $1.96$\\
NGC 2403 \cite{dors2024cosmic} & $80\times 10^{8}$ & $0.09$ & $31.26$ & $6.17$ & $111.05$ & $230.36$ & $47\times 10^{8}$ & $0.03$ & $27.22$ & $7.33$ & $59.61$ & $0.89$\\
NGC 2903 \cite{Silk2002} & $11\times 10^{9}$& $0.09$ & $33.81$& $3.84$ & $14.58$ & $97.50$ & $14\times 10^{9}$& $0.12$ & $33.81$ & $3.90$ & $10.31$ & $0.37$\\
NGC 2998 \cite{chattopadhyay2006objective}  & $60\times 10^{9}$ & $0.03$ & $176.22$& $16.79$ & $36.92$& $344.71$ & $65\times 10^{9}$ & $ 0.21$ & $229.82$ & $20.79$ & $8.33$ & $1.19$\\
NGC 3109 \cite{garling2024dual}&  $4\times 10^{9}$ & $0.11$ & $48.76$ & $7.88$ & $21.92$ & $26.14$ & $4\times 10^{9}$ & $0.11$ & $48.76$ & $7.88$ & $12.24$ & $0.64$\\
NGC 3198 \cite{Reste:2023zpq} & $7\times 10^{9}$  & $0.12$& $21.03$ & $3.83$ & $52.90$ & $59.37$ & $8\times 10^{9}$ & $0.11$ & $18.05$ & $3.22$ & $9.65$ & $0.26$\\
NGC 7331 \cite{Smith2021} & $55\times 10^{8}$& $0.37$& $9.45$&$3.54$ &$7.49$ &$108.24$ & $18\times 10^{8}$ & $0.56$ & $10.45$ & $3.81$ & $15.44$ & $0.51$\\
\hline
\end{tabular}
\label{table1_subhalos_2}
\end{table*}

\begin{figure*}[t]
    \centering
    \includegraphics[width=0.4\textwidth]{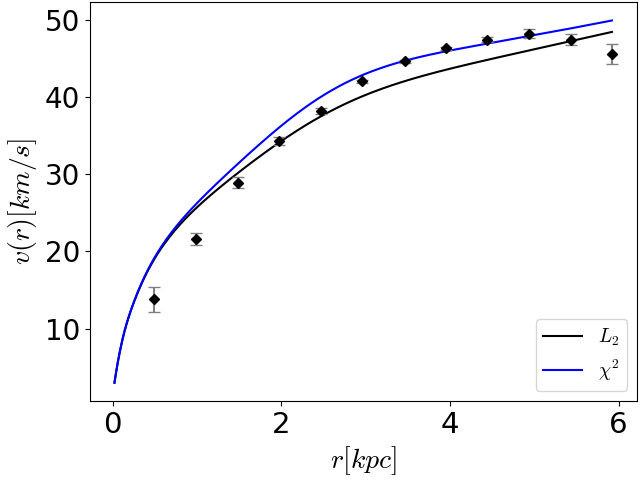}
    \includegraphics[width=0.4\textwidth]{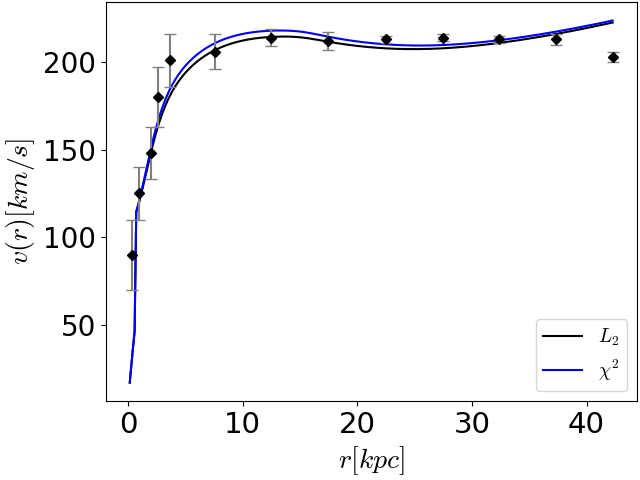}\label{fig:f3}
    \caption{DDO 1154 and NGC 2998 comparing the fits obtained by minimizing the $\chi^2$ (blue) and the $L_2$ (black). Even in the most different cases (shown in this figure), both fits remain comparable. }
    \label{galaxy5}
\end{figure*}

As shown in \cref{table1_subhalos_2}, most values of $\chi^2_{red}$, are close to unity, indicating that the fits are statistically sound. Although the $N\times M_N$ that minimize the $L_2$ norm are not exactly the same as those that minimize $\chi^2$, they are, in general, very similar. In some cases, they even coincide exactly. For those configurations where the values differ more noticeably, the resulting variations in the fits are still minimal from a practical standpoint. Overall, as illustrated in \cref{galaxy5}, the models obtained via $\chi^2$ minimization show only slight deviations from those previously presented.

\bibliography{biblio}

\end{document}